\documentclass{PoS}

\usepackage{amsmath}
\usepackage{graphicx}
\usepackage{bm}

\title{Phase structure and Hosotani mechanism in QCD-like theory with compact dimensions}

\ShortTitle{Phase structure and Hosotani mechanism in QCD-like theory with compact dimensions}

\author{\speaker{Kouji Kashiwa}\thanks{K.K is supported by RIKEN Special
Postdoctoral Researchers Program.}\\
        RIKEN/BNL, Brookhaven National Laboratory, Upton, NY 11973\\
        E-mail: \email{kashiwa@ribf.riken.jp}}

\author{Hiroaki Kouno\\
        Department of Physics, Saga University, Saga 840-8502, Japan\\
        E-mail: \email{kounoh@cc.saga-u.ac.jp}}

\author{Takahiro Makiyama\\
        Department of Physics, Saga University, Saga 840-8502, Japan\\
        E-mail: \email{12634019@edu.cc.saga-u.ac.jp}}

\author{Tatsuhiro Misumi\\
        Keio University, Hiyoshi 4-1-1, Yokohama, Kanagawa 223-8521,
        Japan\\
        E-mail: \email{misumi@phys-h.keio.ac.jp}}

\author{Takahiro Sasaki\\
        Department of Physics, Graduate School of Sciences, Kyushu
        University, Fukuoka 812-8581, Japan\\
        E-mail: \email{sasaki@phys.kyushu-u.ac.jp}}

\author{Masanobu Yahiro\\
        Department of Physics, Graduate School of Sciences, Kyushu
        University, Fukuoka 812-8581, Japan\\
        E-mail: \email{yahiro@phys.kyushu-u.ac.jp}}

\abstract{
We investigated the phase diagram of $SU(3)$ gauge theory in
four dimension with one compact dimension by using
the perturbative one-loop effective potential.
Effects of the adjoint and fundamental fermions are investigated and then
the rich phase structure in the quark-mass and compact-size scale is
realized.
Our results are qualitatively consistent with the recent lattice
calculation and clearly show that the lattice calculation can be understood
from the Hosotani mechanism.
Moreover, we show the result obtained by using the flavor twisted
boundary condition for fundamental fermion which does not break the
$Z_3$ symmetry, explicitly.
}

\FullConference{31st International Symposium on Lattice Field Theory - LATTICE 2013\\
		July 29 - August 3, 2013\\
		Mainz, Germany}

\begin{document}

\section{Introduction}

The Higgs-like particle has been discovered recently in Large Hadron
Collider (LHC) \cite{Aad:2012tfa,Chatrchyan:2012ufa}.
The one of primary interests of particle physics is to understand the mechanism of
dynamical electroweak symmetry breaking.
The one of the promising mechanism to explain the Higgs particle is the
Hosotani mechanism~\cite{Hosotani:1983xw,Hosotani:1988bm} which leads
the gauge-Higgs unification.

In the Hosotani mechanism, the Higgs particle is interpreted as the
fluctuation of the extra-dimensional component of the gauge field when
the adjoint fermions are introduced with a periodic boundary condition
(PBC) because the non-zero vacuum expectation value (VEV) of the
extra-dimensional component of gauge field is realized.

Recently, same phenomena has been observed in a different context
; for example, see Ref.~\cite{Nishimura:2009me,Cossu:2009sq,Cossu:2013ora}.
When the adjoint fermions with PBC are introduced to Quantum
Chromodynamic (QCD) at finite temperature, some exotic phases are
appeared.
In such exotic phase, the traced fundamental Polyakov-loop $\Phi$
can have the non-trivial value and it show the spontaneous
gauge-symmetry breaking.
It means the realization of the Hosotani mechanism in $R^3 \times S^{1}$
space-time as shown later.

Furthermore, we consider the flavor twisted boundary condition (FTBC) for
fundamental fermions.
This FTBC is considered in Ref.~\cite{Kouno:2012zz,Sakai:2012ika} to
investigate correlations between the $Z_3$ and chiral symmetries
breaking because the $Z_3$
symmetry is not explicitly broken in the case with FTBC
even if we introduce the fundamental fermions.
In the standard fundamental fermion can not leads the spontaneous gauge
symmetry breaking, but fundamental fermions with FTBC can lead the
breaking as shown later.

The purpose of this talk is to explain
how to understand the recent lattice simulation
from the Hosotani mechanism and possibility of the spontaneous
gauge symmetry breaking by the fundamental fermions.
This talk is based on papers~\cite{Kashiwa:2013rmg,Kouno:2013mma}

\section{Formalism}

In this study, we use the perturbative one-loop effective potential
~\cite{Gross:1980br,Weiss:1980rj}
on $R^{3}\times S^1$ for gauge boson and fermions and then the imaginary
time direction is the compacted dimension.
%When we consider the chiral symmetry breaking and restoration, the
%four-fermi and eight-fermi interactions are introduced.

Firstly, we expand the $SU(N)$ gauge boson field as
\begin{align}
A_\mu &= \langle A_y \rangle + \tilde{A}_\mu,
\end{align}
where $y$ stands for a compact direction, $\langle A_y \rangle$
is VEV and $\tilde{A}_\mu$ express the fluctuation part.
For latter convenience, we rewrite it as
\begin{align}
\langle A_y \rangle &=\frac{2 \pi}{gL} q,
\end{align}
where $g$ is gauge coupling constant and
$q$'s color structure is $\mathrm{diag}(q_1,q_2,...,q_{N})$
and each component should be $(q_i)_{mod~1}$.
We note that eigenvalues of $q_{i}$
are invariant under all gauge transformations preserving boundary conditions
and thus we can easily observe spontaneous gauge symmetry breaking from
values of $q_{i}$.

The gluon one-loop effective potential ${\cal V}_g$ can be expressed as
\begin{align}
{\cal V}_g
&= - \frac{2}{L^4 \pi^2} \sum_{i,j=1}^N \sum_{n=1}^{\infty}
     \Bigl( 1 - \frac{1}{N} \delta_{ij} \Bigr)
     \frac{\cos( 2 n \pi q^{ij})}{n^4}
\end{align}
where $q_{ij} = ( q_i - q_j )_{mod~1}$
and $N$ means the number of color degrees of freedom.
The perturbative one-loop effective potential for the massive fundamental quark
is expressed by using the second kind of the modified Bessel function $K_2(x)$ as
\begin{align}
{\cal V}_f^\phi(N_{f},m_f) &=
\frac{ 2 N_f m_{f}^{2}}{\pi^2L^2} \sum_{i=1}^N \sum_{n=1}^\infty 
\frac{K_2 ( n m_{f} L )}{n^2}
\cos [2 \pi n (q_i + \phi)],
\end{align}
where $N_f$ and $m_{f}$ are the number of flavors and the mass for fundamental fermions.
The perturbative one-loop effective potential for the massive adjoint
quark ${\cal V}_a^\phi$ is
\begin{align}
{\cal V}_a^\phi (N_{a}, m_{a}) &=
\frac{ 2 N_a m_{a}^{2}}{\pi^2L^2} \sum_{i,j=1}^N \sum_{n=1}^\infty
\Bigl(  1 - \frac{1}{N} \delta_{ij} \Bigr)
\frac{K_2 ( n m_{a} L )}{n^2}
\cos [2 \pi n (q_{ij} + \phi)],
\label{Re}
\end{align}
where $N_a$ and $m_{a}$ are the number of flavors and the mass for adjoint fermions.

For the gauge theory with $N_{f}$ fundamental and $N_{a}$ adjoint fermions with
arbitrary boundary conditions,
the total perturbative one-loop effective potential becomes
\begin{equation}
{\cal V} = {\cal V}_{g}+{\cal V}_{f}^{\phi}(N_{f}, m_f)+{\cal V}_a^{\phi}(N_{a}, m_a).
\label{eq_ep_pert}
\end{equation}
This total one-loop effective potential contains eight parameters including
the compact scale $L$, the number of colors $N$, the fermion
masses $m_{f}$, $m_{a}$, the number of flavors $N_{f}$, $N_{a}$, and the
boundary conditions $\phi$ for two kinds of matter fields.
In this study, we keep $N=3$ and then the phase diagram is obtained in
$1/L$-$m_{a}$ space with fixed $m_{f}$, $N_{f}$, $N_{a}$ and $\phi$.
The reason we change $m_{a}$ while fixing $m_{f}$ is that gauge symmetry
phase diagram is more sensitive to the former than the latter.

\section{ $SU(3)$ gauge theory with adjoint and fundamental
 quarks~\cite{Kashiwa:2013rmg}}

\begin{figure}[htbp]%[H]
\begin{center}
 \includegraphics[width=0.23\textwidth]{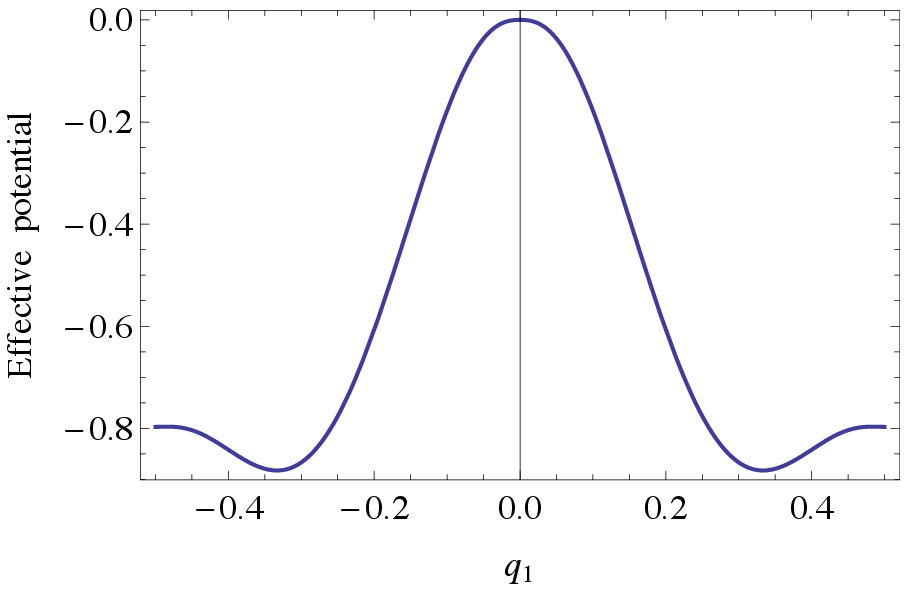}
 \includegraphics[width=0.23\textwidth]{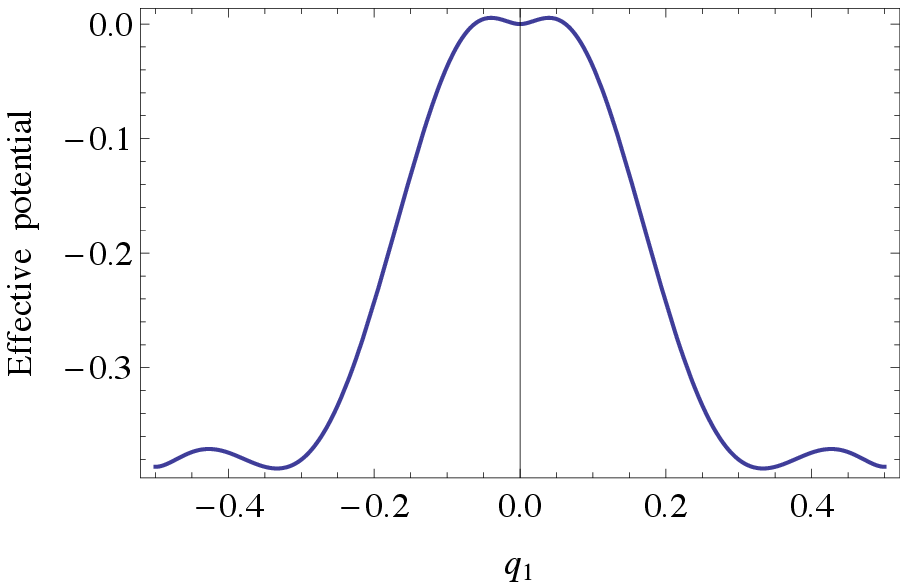}
 \includegraphics[width=0.23\textwidth]{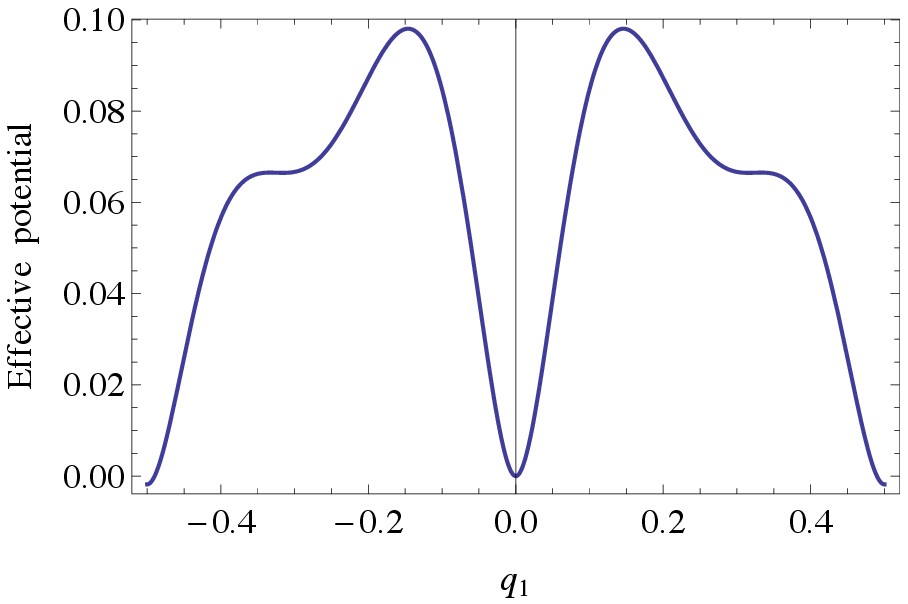}
 \includegraphics[width=0.23\textwidth]{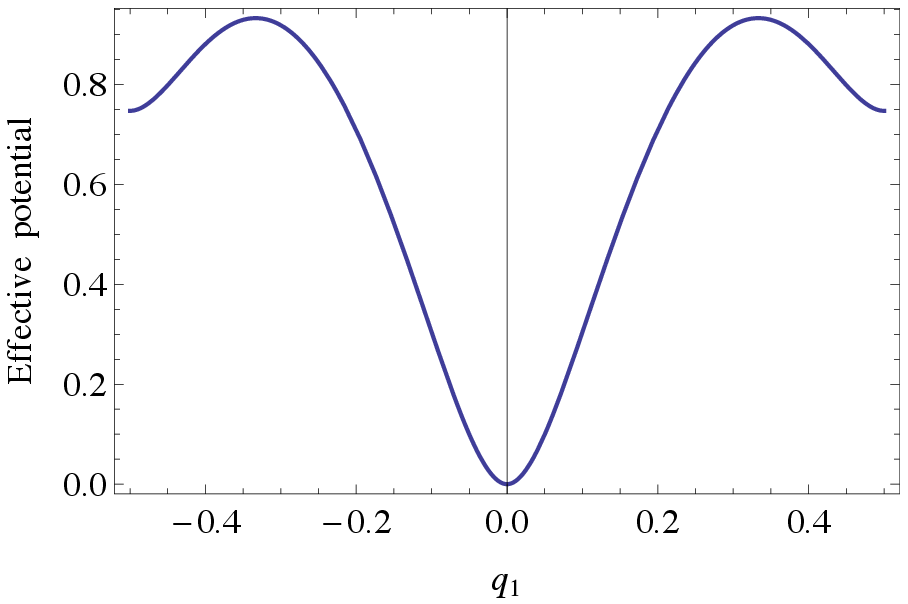}
\end{center}
\caption{ 
The one-loop effective potential of $SU(3)$ gauge theory
with one flavor PBC adjoint quark as a function of $q_1$ with $q_2=0$
for $m L=1.2$ (reconfined)
$1.6$ (reconfined$\leftrightarrow$split),
$2.0$ (split$\leftrightarrow$deconfined) and $3.0$ (deconfined).
}
\label{Fig_p_gapm_2D}
\end{figure}

Here, we consider the case of $(N_{f},N_{a})=(0,1)$ with PBC.
We note that this case has exact $Z_3$ symmetry because the adjoint
quark dose not break the symmetry.
Figure~\ref{Fig_p_gapm_2D} shows the effective potential
$[{\cal V}_g+{\cal V}_a^{0}(N_{a}, m_{a})]L^4$ as a function of $q_{1}$
with $q_{2}=0$ for $m L=1.2$, $1.6$, $2.0$ and $3.0$ from left to right
panels ($m \equiv m_{a}$).
%%%%%%%%%%%%%%%% Fig %%
%%%%%%%%%%%%%%%%%%%%
We can clearly see that there is the first-order phase transition in
the vicinity of $m L=1.6$.
This is a transition between the reconfined phase and the other gauge-broken
phase, which we call the split phase.
%The contour plots for $mL=1.6$ and $mL=1.8$ are shown in
%Fig.~\ref{Fig_p_gapm_3D}.
%The $mL=1.8$ corresponds to the split phase.
%%%%%%%%%%%%%%%% Fig %%
%\begin{figure}[htbp]%[H]
%\begin{center}
% \includegraphics[width=0.4\textwidth]{potential_gap_m16p05_CP.eps}
% \includegraphics[width=0.4\textwidth]{potential_gap_m18_CP.eps}
%\end{center}
%\caption{
%Contour plot of the perturbative one-loop effective potential of $SU(3)$
%gauge theory with one flavor PBC adjoint quark
%$[ ( {\cal V}_g )_{pert} + {\cal V}_a^{0} ] L^4$,
%for $m L=1.6$ and $1.8$ ($SU(2)\times U(1)$ split phase)
%as a function of $q_1$ and $q_2$.
%Thicker region indicates deeper region of the potential.}
%\label{Fig_p_gapm_3D}
%\end{figure}
%%%%%%%%%%%%%%%%%%%%

\begin{figure}[htbp]%[H]
\begin{center}
 \includegraphics[width=0.55\textwidth]{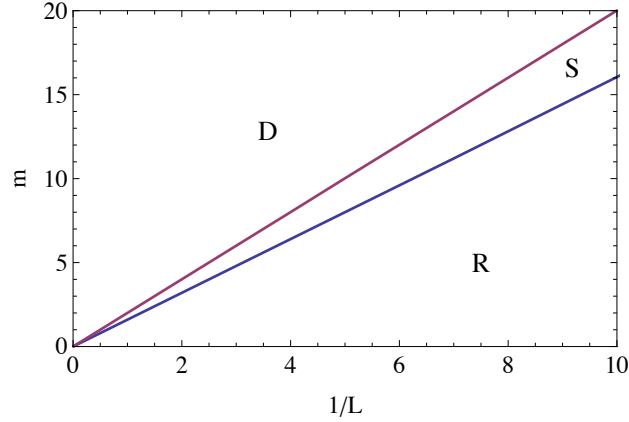}
\end{center}
\caption{$L^{-1}$-$m$ phase diagram for $SU(3)$ gauge theory on 
$R^{3}\times S^1$ with one PBC adjoint quark
based on one-loop effective potential.
The symbol D stands for deconfined ($SU(3)$),
S for split ($SU(2)\times U(1)$) and R for reconfined ($U(1)\times U(1)$) phases.}
\label{4d_phase_p}
\end{figure}

\begin{figure}[htbp]%[H]
\begin{center}
 \includegraphics[width=0.45\textwidth]{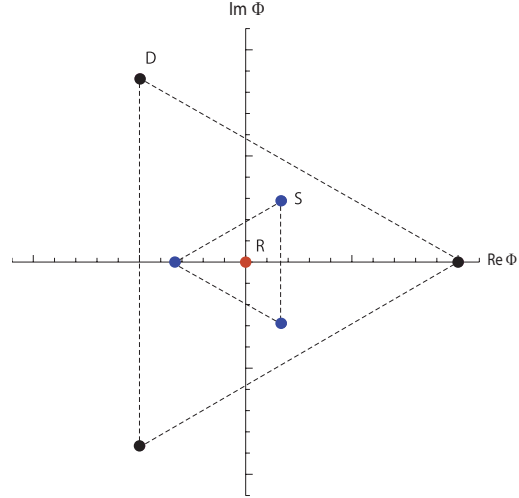}
\end{center}
\caption{
Schematic distribution plot of Polyakov loop $\Phi$ as a function of
$\mathrm{Re}~\Phi$ and $\mathrm{Im}~\Phi$ for $SU(3)$ gauge theory
with one flavor PBC adjoint quark.
}
\label{Fig_pl_distribution}
\end{figure}

In Fig.~\ref{4d_phase_p}, we show the phase diagram in $L^{-1}$-$m$ plane
with $(N_{f}, N_{a})=(0,1)$ quark based on the perturbative one-loop
effective potential.
%%%%%%%%%%%%%%%% Fig %%
%%%%%%%%%%%%%%%%%%%%
We note that, as $m$ appears as $m L$ in the potential,
we have liner scaling in the phase diagram.
Since we drop the non-perturbative effect in the gluon potential,
we can not obtain the confined phase at small $L^{-1}$.
The order of three phases in Fig.~\ref{4d_phase_p}
(deconfined $SU(3)$ $\to$ split $SU(2)\times U(1)$ $\to$ reconfined $U(1)\times U(1)$
from small to large $L^{-1}$)
is consistent with that of the lattice simulation~\cite{Cossu:2009sq,Cossu:2013ora}
except the confined phase.
All the critical lines in the figure are first-order.
In Fig.~\ref{Fig_pl_distribution} we show a schematic distribution
plot of $\Phi$ in the complex plane for each phases.
In the split phase, $\Phi$ has nonzero values but in a different manner from the deconfined phase.
In the reconfined phase, we have $\Phi=0$ with the vacuum which breaks the gauge symmetry.
%%%%%%%%%%%%%%%% Fig %%
%%%%%%%%%%%%%%%%%%%%

\begin{figure}[htbp]%[H]
\begin{center}
 \includegraphics[width=0.65\textwidth]{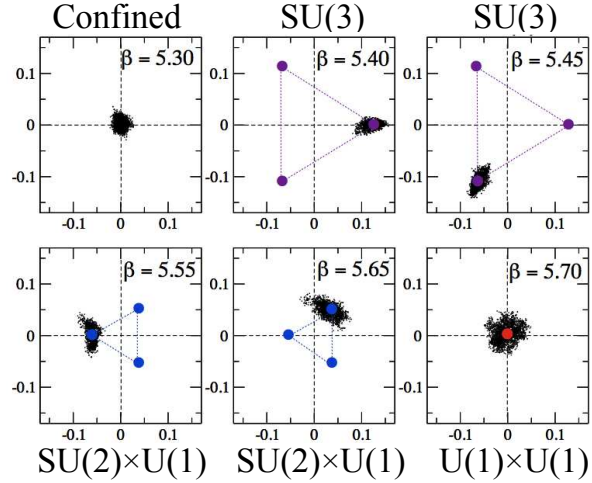}
\end{center}
\caption{Comparison between distribution plot of Polyakov loop $\Phi$ on the lattice \cite{Cossu:2009sq} 
and that of the one-loop effective potential for $SU(3)$ gauge theory on 
$R^{3}\times S^1$ with PBC adjoint quarks. Apart from the strong-coupling confined phase,
all of the specific behavior can be interpreted as the phases we found in our analytical calculations. }
\label{PoD}
\end{figure}

From above results, we can understand the lattice results~\cite{Cossu:2009sq}
from Hosotani mechanism as shown in Fig~\ref{PoD}.
%%%%%%%%%%%%%%%% Fig %%
%%%%%%%%%%%%%%%%%%%%

\begin{figure}[htbp]%[H]
\begin{center}
\includegraphics[bb=0 150 850 500, clip, width=0.9\textwidth]{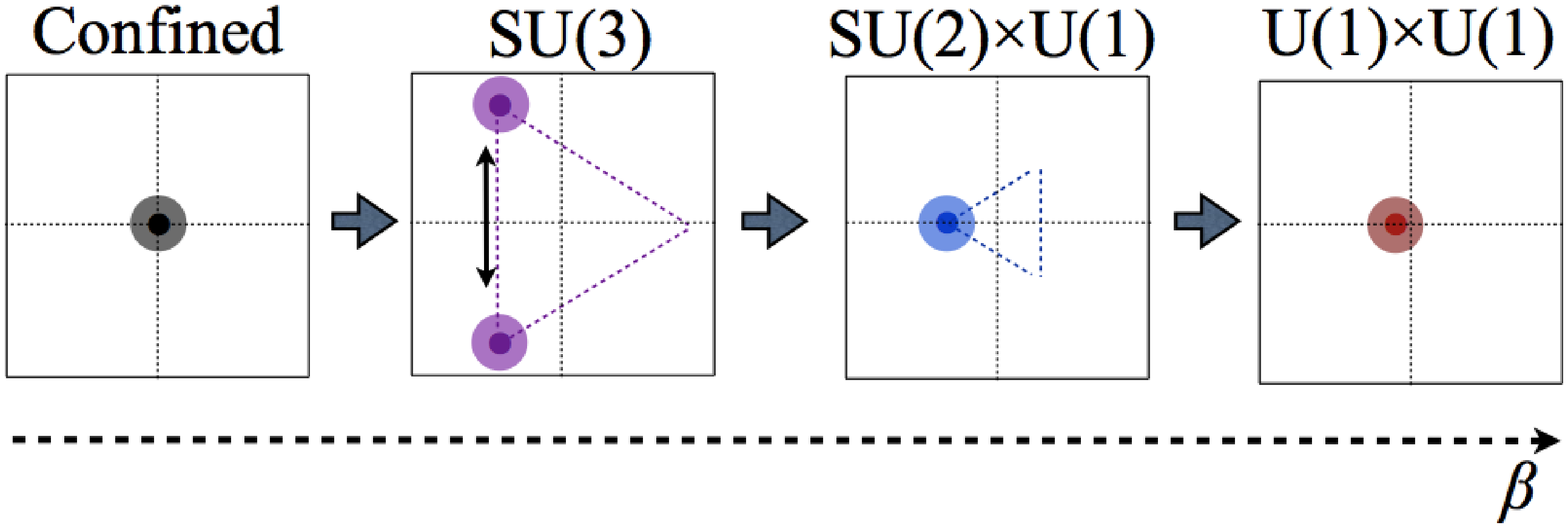} 
\end{center}
\caption{Prediction of distribution plot of Polyakov loop $\Phi$
based on the one-loop effective potential for $SU(3)$ gauge theory on 
$R^{3}\times S^1$ with PBC adjoint and fundamental quarks.}
\label{PoD2}
\end{figure}

The schematic figure of the fundamental quark effect to the phase
diagram is shown in Fig.~\ref{PoD2}.
%%%%%%%%%%%%%%%% Fig %%
%%%%%%%%%%%%%%%%%%%%
The reconfined phase is replaced by the pseudo-confined phase because
the $Z_3$ symmetry is explicitly broken by the fundamental quark
contributions,
but the gauge symmetry breaking pattern is still same.

\section{ $SU(3)$ gauge theory with FTBC fundamental
 quarks~\cite{Kouno:2013mma}}

In this section, we consider the FTBC for the fundamental fermion.
Details of FTBC are shown in Ref.~\cite{Kouno:2012zz,Sakai:2012ika}.

\begin{figure}[htbp]%[H]
\begin{center}
 \includegraphics[width=0.35\textwidth]{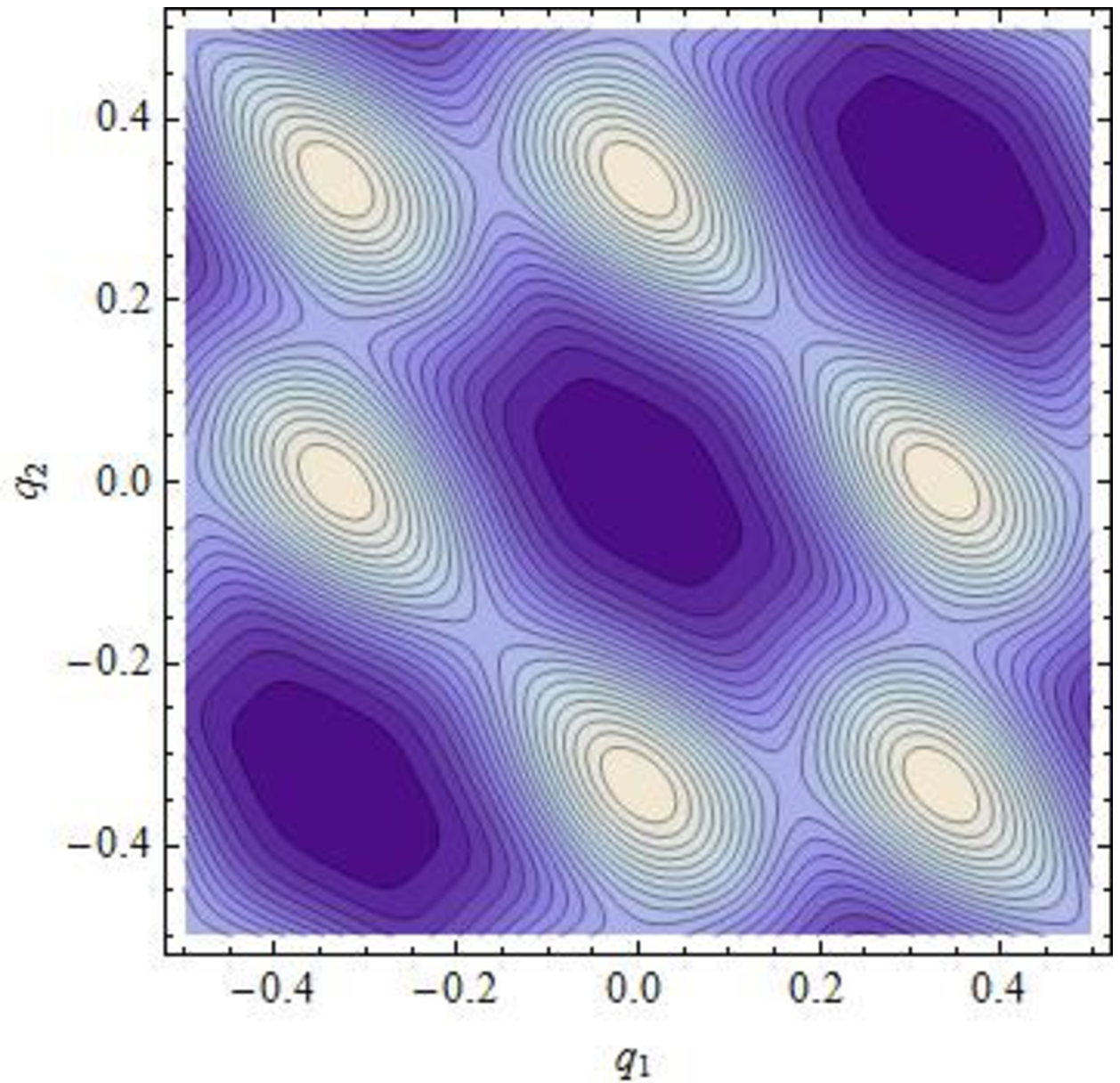}
 \includegraphics[width=0.35\textwidth]{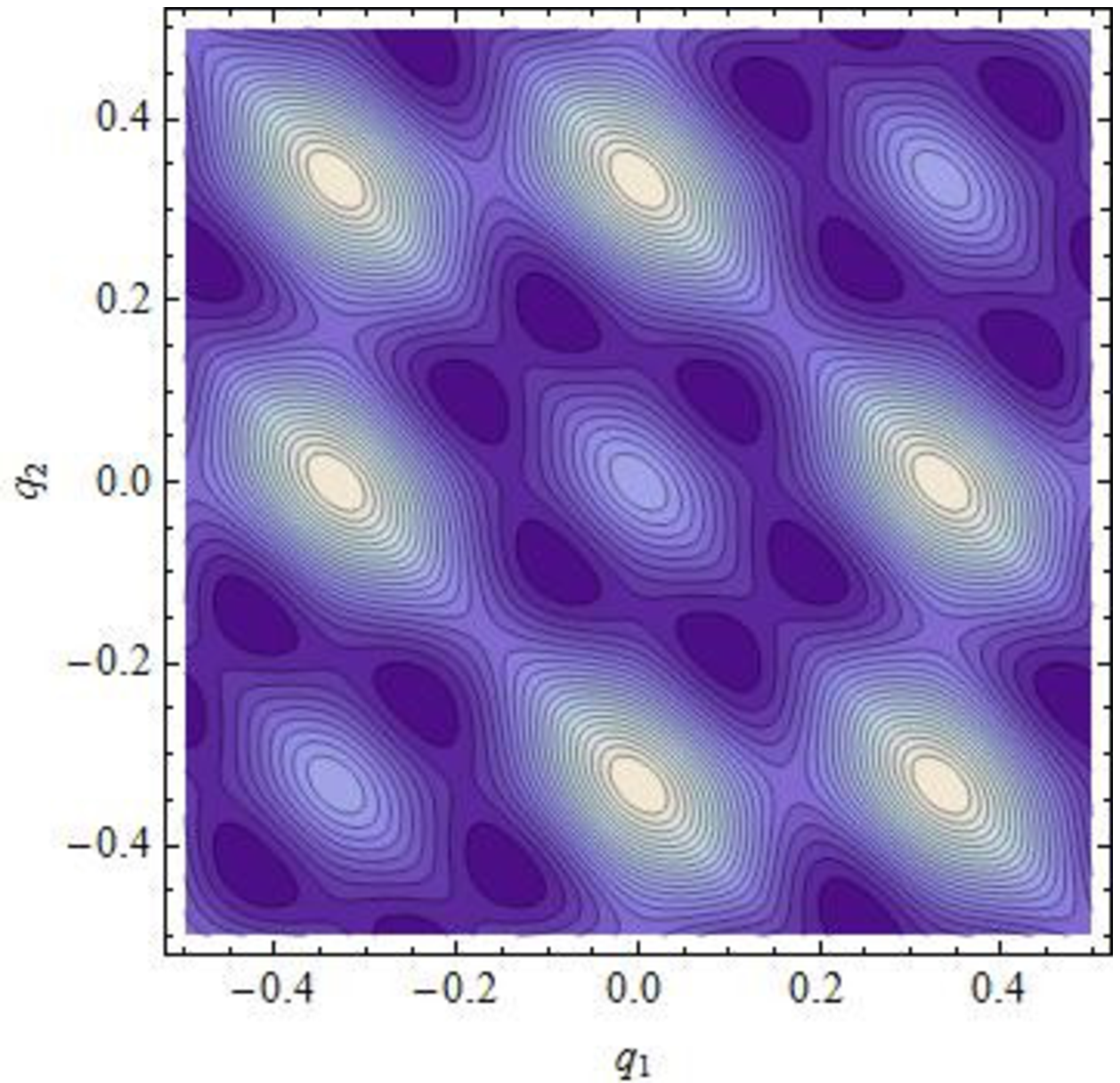}
\end{center}
\caption{
Contour plot of $[  {\cal V}_{g} + {\cal V}_{f} ] L^4$ 
in the  $q_1$-$q_2$ plane for the case of $N_{F,fund}=120$ FTBC fermions. 
The upper panel corresponds to the $SU(3)$ deconfined phase and 
the lower panel does to the $SU(2)\times U(1)$ C-broken phase.}
\label{CP}
\end{figure}

\begin{figure}[htbp]%[H]
\begin{center}
 \includegraphics[width=0.6\textwidth]{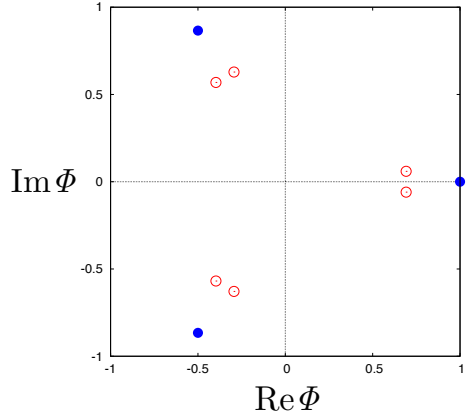}
\end{center}
\caption{
Distribution of the Polyakov loop in the complex plane for a $SU(3)$ gauge theory on
with $N_{F,fund}=120$ FTBC fermions.
Solid circles correspond to the deconfinement phase and open circles do
 to the gauge symmetry broken phase.
}
\label{DP}
\end{figure}

Contour plots of the $SU(3)$ gauge theory with $N_{F,fund}=120$ FTBC
fundamental quark for the gauge symmetric and broken phase are shown
in Fig.~\ref{CP}.
%%%%%%%%%%%%%%%% Fig %%
%%%%%%%%%%%%%%%%%%%%
Unlike the standard fundamental quark, we can clearly see the existence
of the spontaneous gauge symmetry breaking of $SU(3) \to SU(2) \times U(1)$.
The distribution plot of the fundamental Polyakov-loop is shown in Fig.~\ref{DP}.
%%%%%%%%%%%%%%%%% Fig %%
%%%%%%%%%%%%%%%%%%%%

\begin{figure}[htbp]%[H]
\begin{center}
 \includegraphics[width=0.5\textwidth]{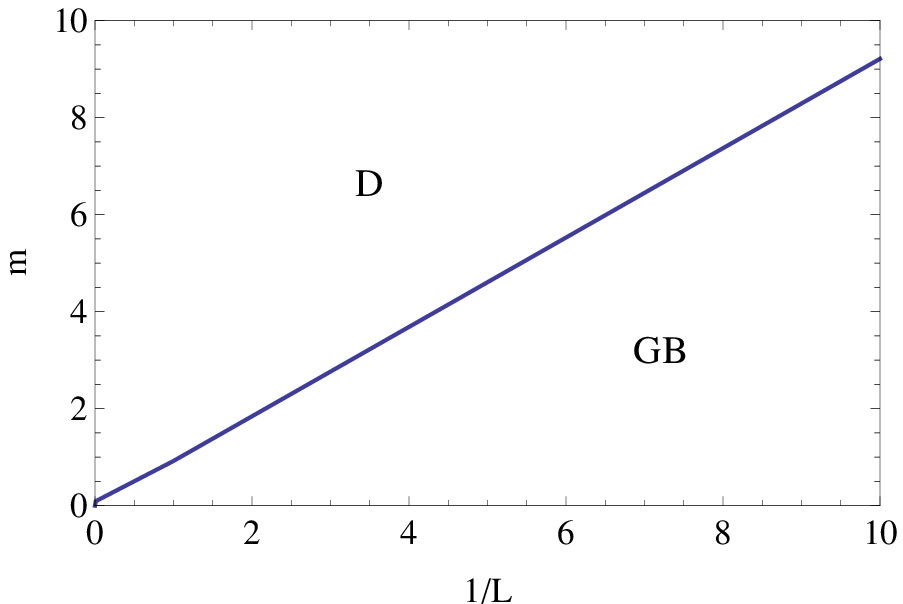}
\end{center}
\caption{The phase diagram in the $L^{-1}$-$m$ plane
for a $SU(3)$ gauge theory with $N_{F,fund}=120$ FTBC fermions.
The symbol D stands for the deconfinement phase
and GB for the $SU(2)\times U(1)$ gauge symmetry broken phase.
In the gauge symmetry broken phase, charge conjugation is also
 spontaneously broken which can be seen from the charge-conjugation pairs.}
\label{PD}
\end{figure}

The phase diagram is shown in Fig.~\ref{PD} in the $L^{-1}$-$m$ plane.
%%%%%%%%%%%%%%%% Fig %%
%%%%%%%%%%%%%%%%%%%%
In the case with FTBC fundamental fermions, there is no $U(1) \times
U(1)$ phase, but $SU(2) \times U(1)$ phase is still exist.
The $Z_3$ symmetry is not explicitly broken as same as the adjoint
fermions
and also lattice simulations are possible.
Therefore, this system is very interesting to consider the
gauge symmetry breaking and also the confinement-deconfinement transition.

\end{document}